\documentclass[twocolumn,showpacs,amsmath,amssymb,pr,superscriptaddress]{revtex4-1}
\usepackage{graphicx}
\usepackage{ulem}
\usepackage{color}
\usepackage[unicode=true,colorlinks=true,urlcolor=blue,citecolor=blue]{hyperref}

\begin{document}

\title{Intersubband scattering in n-GaAs/AlGaAs wide quantum wells}

\author{I.~L.~Drichko}
\author{I.~Yu.~Smirnov}
\author{M.~O.~Nestoklon}
\affiliation{Ioffe Institute, 194021 St. Petersburg, Russia}
\author{A.~V.~Suslov}
\affiliation{National High Magnetic Field Laboratory, Tallahassee, FL 32310, USA}
\author{D.~Kamburov}
\author{K.~W.~Baldwin}
\author{L.~N.~Pfeiffer}
\author{K.~W.~West}
\affiliation{Department of Electrical Engineering, Princeton University, Princeton, NJ 08544, USA}
\author{L.~E.~Golub}
\affiliation{Ioffe Institute, 194021 St. Petersburg, Russia}

\begin{abstract}
Slow magnetooscilations of the conductivity are observed in a 75 nm wide quantum well at heating of the two-dimensional electrons by a high-intensity surface acoustic wave. These magnetooscillations are caused by intersubband elastic scattering between the symmetric and asymmetric subbands formed due to an electrostatic barrier in the center of the quantum well. The tunneling splitting between these subbands as well as the intersubband scattering rate are determined.
\end{abstract}

\pacs{73.63.Hs, 73.50.Rb}

\date{\today}

\maketitle

\section{Introduction} \label{introduction}

Quantum structures with more than one occupied levels of size quantization represent an intermediate case between ultra-quantum and bulk systems. A presence of a few two-dimensional subbands allows studying interactions between electronic states of different types. An interesting example is intersubband scattering by a disorder potential. The typical systems with a few levels are quantum wells with two or more subbands under the Fermi level and double quantum wells. There is also another type of structures, doped wide quantum wells (WQWs). They represent a bilayer system because the Coulomb repulsion results in a potential barrier in the middle of the WQW pushing the carriers towards the interfaces~\cite{Shayegan_120nm}.
If these two layers are independent, they act in transport as two parallel conducting channels. These two channels are identical with equal Fermi energies and relaxation times provided the WQW is perfectly symmetric.
In contrast, when tunneling through the potential barrier is not negligible, these two channels interact to each other, and the system's eigenstates are the symmetric (S) and anti-symmetric (AS) states with the tunneling energy gap $\Delta_\text{SAS}$. This gap has been studied in a variety of WQWs, for a review see Ref.~\cite{shayegan_review}.
Usually $\Delta_\text{SAS}$ is determined from the Fourier analysis of the magnetoresistance in the region of weak magnetic fields $B<0.5$~T.

The presence of two channels results in a reach picture of conductivity oscillations in quantizing magnetic fields. In addition to the usual  Shubnikov-de~Haas effect, the other type of magnetooscillations periodic in $1/B$ takes place.
These oscillations are caused by elastic scattering between the S and AS subbands, the so-called magneto-intersubband oscillations (MISO).
They appear at $\Delta_\text{SAS}/\hbar\omega_c = K$, where $\omega_c$ is the cyclotron frequency and $K$ is an integer number.
Since this condition does not contain the Fermi energy, MISO are not damped by the Fermi distribution smearing. Therefore, in contrast to the Shubnikov-de~Haas oscillations, MISO amplitude is almost insensitive to the temperature increase.
MISO are well studied in various systems with two or three occupied subbands, for a review see Ref.~\cite{I_Dmitriev_review} and references therein. Recently, a temperature dependence of MISO amplitude in a quantum well with three populated subbands has been explained by temperature variation of quantum electron lifetime~\cite{tau_q_3_subbands}, an energy spectrum reconstruction by a parallel magnetic field has been shown to affect MISO strongly~\cite{MISO_tilted_field_1,MISO_tilted_field_2}, and the thermoelectric power magnetophonon resonance has been studied in two-subband quantum wells~\cite{Thermopower_m_osc}.

MISO are possible to observe only if they are not superimposed on the Shubnikov-de~Haas oscillations. However both types of oscillations are present in the same magnetic field range in high-mobility WQWs.
The Shubnikov-de~Haas oscillations can be damped by increase of temperature.
However, heating of the sample in dc regime also results in an increase of the lattice temperature. This leads to an enhancement of electron scattering by  phonons which damps MISO as well.
Therefore MISO in high-mobility WQWs have not been observed so far.

We used acoustic methods with a surface
acoustic wave (SAW) of high intensity applied in the pulsed regime with the duty factor equal to 100.
 This allowed heating  of the electron system  up to $T > 500$~mK while the lattice temperature was kept 20~mK. As a result, the Shubnikov-de~Haas oscillations were damped, and clear MISO were observed.
We analyzed MISO in WQWs and determined the energy gap $\Delta_\text{SAS}$ and the intersubband scattering rate.
We show that the theory of magnetooscillations describes well the experimental data.

\section{Experiment} \label{experimental}

The high quality samples were multilayer n-GaAlAs/GaAs/GaAlAs
structures with a 75~nm wide quantum well. The quantum GaAs well
was $\delta$-doped on both sides and located at the depth
$\approx 197$~nm below the surface of the sample. While cooling the sample down to 15~K and illuminating it with infrared light of emitting diode, we achieved the electron density of $1.4\times
10^{11}$~cm$^{-2}$ and the mobility of $2.4\times
10^7$~cm$^2$/(Vs) (at $T$=0.3~K).

\begin{figure}[h]
\centering
\includegraphics[width=0.8\columnwidth]{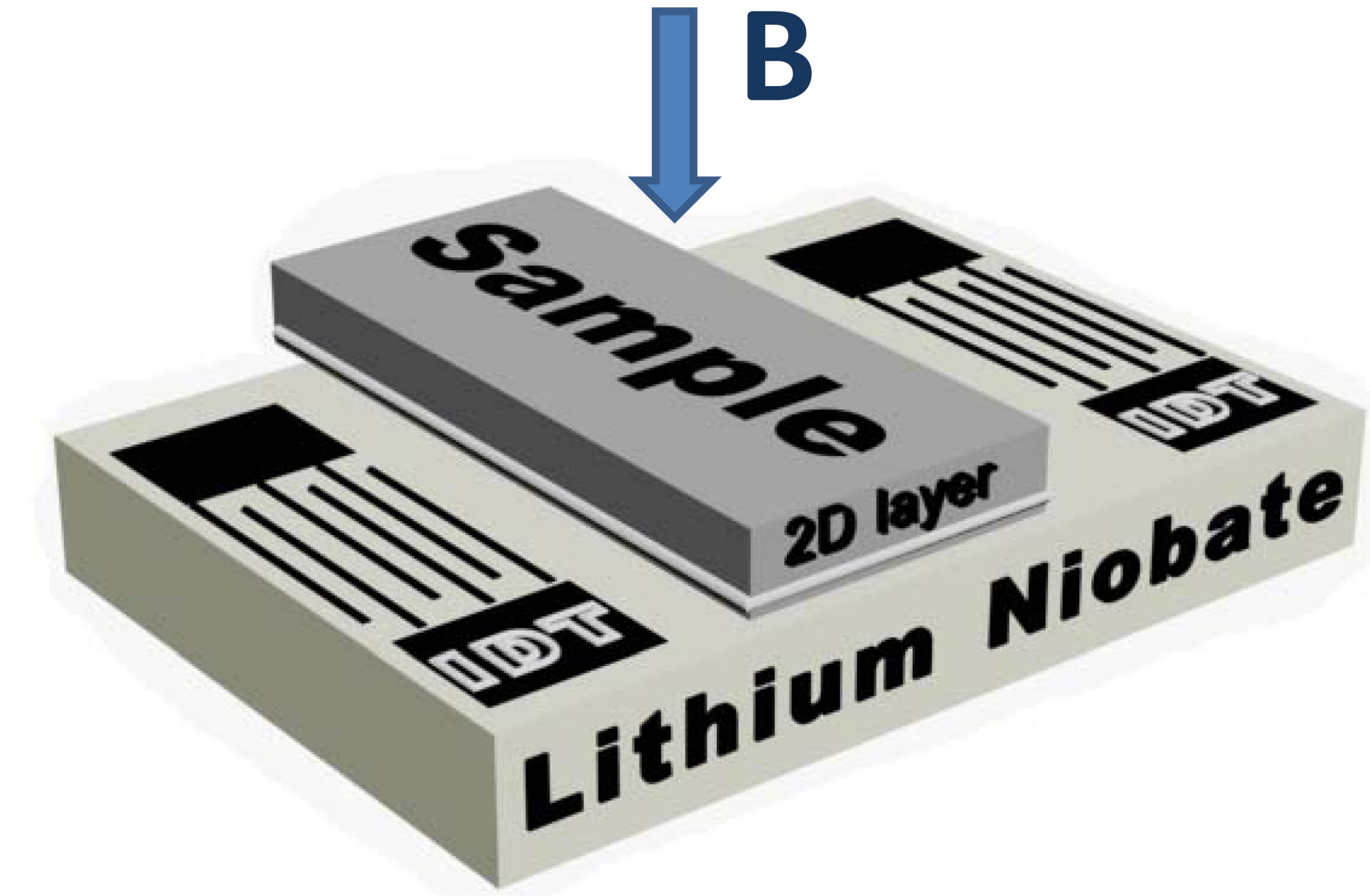}
\caption{(Color online) Sketch of the experimental setup.
 \label{fig1}}
\end{figure}

In the present paper we employ a SAW
technique~\cite{Wixforth1989,Drichko2000} illustrated in
Fig.~\ref{fig1}.
A sample is pressed by means of
springs to the surface of a piezoelectric crystal of lithium niobate
(LiNbO$_3$), on which the interdigitated transducers (IDT) are
formed. A radio frequency electrical pulse
signal is applied to one of the IDTs. Due to the piezoelectric effect, a SAW is generated and propagates along the surface of
LiNbO$_3$. Simultaneously, an ac electric field, accompanying the
SAW and having the same frequency, penetrates into the sample and
interacts with the charge carriers. This interaction results in a
change of the SAW amplitude and in its velocity. The measurements
were carried out in a dilution refrigerator in a magnetic field
perpendicular to the sample plane.

\subsection{Experimental results}

\begin{figure}[h]
\centering
\includegraphics[width=0.8\columnwidth]{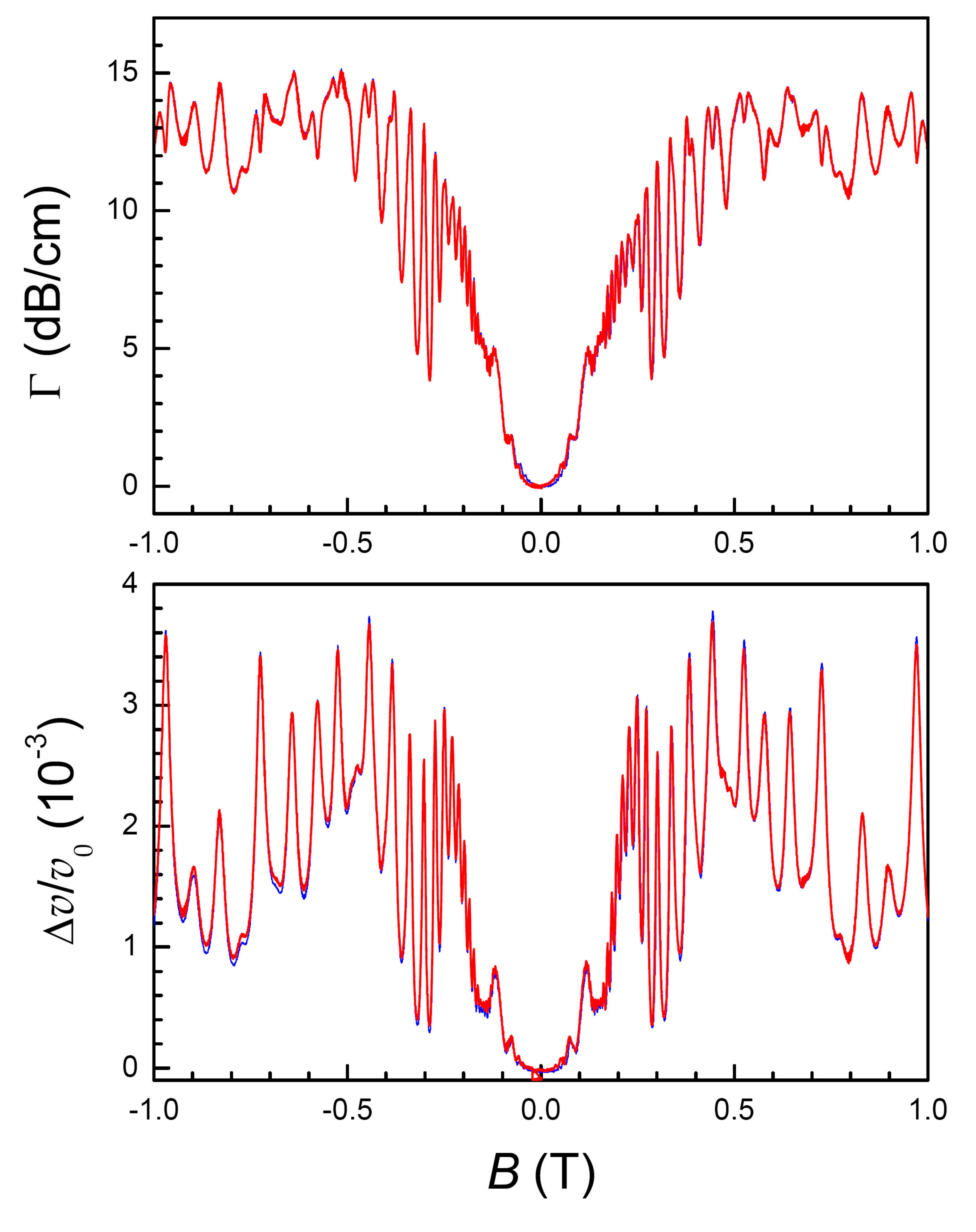}
\caption{(Color online) Dependences of the SAW attenuation coefficient $\Gamma$ (top panel) and the SAW
velocity change $\Delta v (B)/v_0$ (bottom panel) on the transverse magnetic field $B$ at $f$=30~MHz, $T=20$~mK; SAW power introduced into the sample is $1.2\times10^{-6}$~W/cm.
Red and blue curves (almost identical) show forward and reverse field sweeps.
 \label{fig2}}
\end{figure}

The dependences of the attenuation $\Gamma (B)$ and the relative
velocity change $\Delta v (B)/v_0$ of the surface acoustic wave were
measured in a magnetic field of up to 1~T in the
temperature range 20$\div$500~mK and the frequency range
28.5$\div$300~MHz at different SAW intensities. Figure~\ref{fig2}
shows the experimental dependencies of the SAW attenuation
$\Gamma$ and velocity shift $\Delta v/v_0$ at the frequency 30~MHz,
measured at the temperature $T \approx 20$~mK  with the SAW power introduced into the sample of
$1.2\times10^{-6}$~W/cm. During the measurements, the magnetic
field was swept from $-1$ to 1~T (red curve), and then went back to
$-1$~T (blue curve) ramping as 0.05~T/min. The curves of these forward and reverse field sweeps are almost identical. A Hall probe was used to measure the magnetic field strength.

The SAW attenuation and the velocity change are governed by the
complex ac conductance ${\sigma (\omega) \equiv \sigma_1 (\omega)
-i\sigma_2 (\omega)}$.
Both the real $\sigma_1$ and imaginary
$\sigma_2$ components of $\sigma (\omega)$ could be extracted from
our acoustic measurements. The procedure of the determination of the ac conductance is
described in Ref.~\cite{Drichko2000} and is based on using that work Eqs. (1)$\div$(7).

\begin{figure*}[t]
\centering
\includegraphics[width=0.8\textwidth]{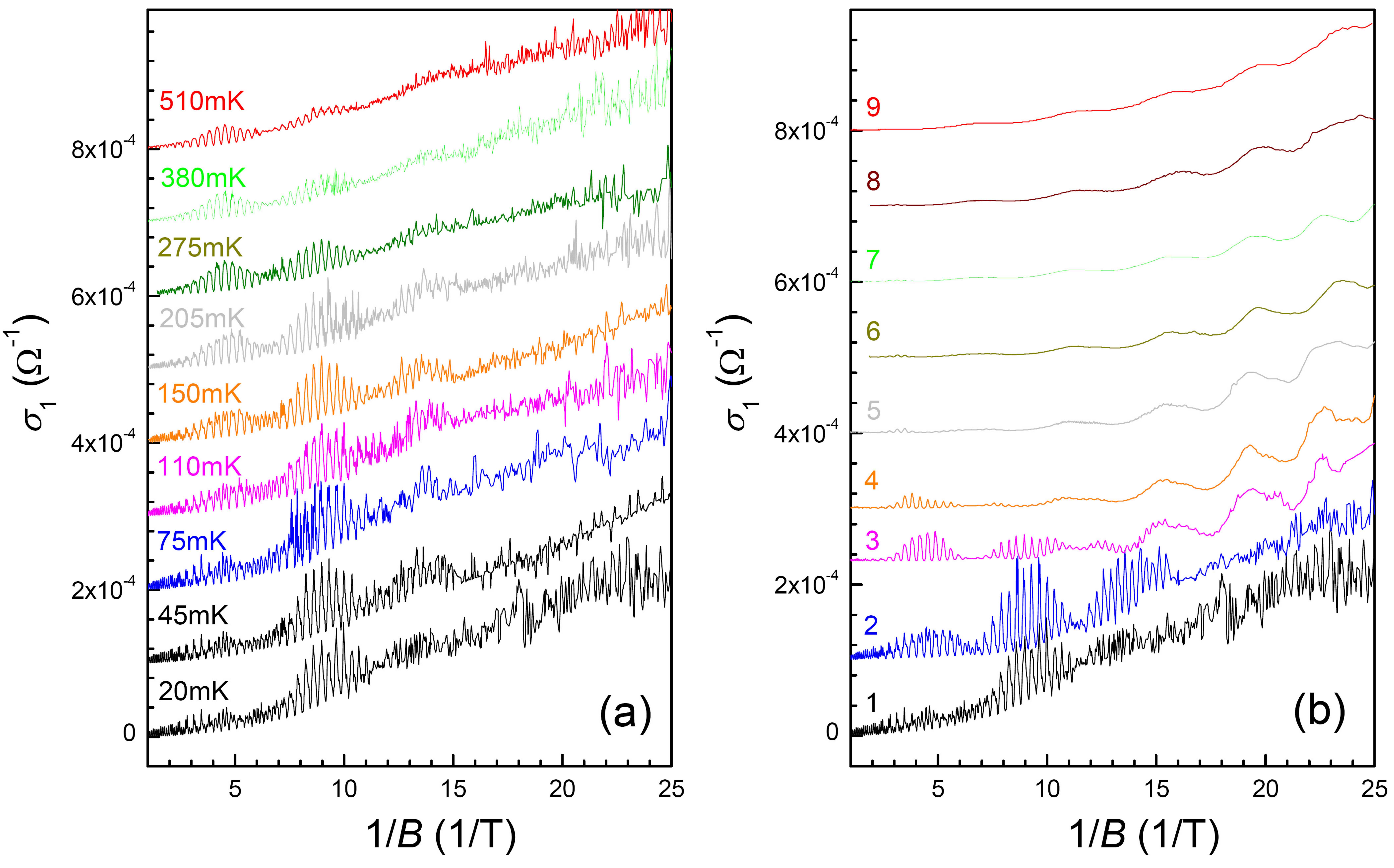}
\caption{(Color online) (a) Dependences of $\sigma_1$ on the inverse magnetic field as
varied with temperature at the SAW power introduced into
the sample of $3.7\times10^{-10}$~W/cm, and (b) as
varied with the SAW
powers at $T$=20~mK: 1 - 3.7$\times$10$^{-10}$~W/cm, 2 -
1.2$\times$10$^{-8}$~W/cm, 3 -  1.3$\times$10$^{-7}$~W/cm, 4 -
3.6$\times$10$^{-7}$~W/cm, 5 - 1.2$\times$10$^{-6}$~W/cm, 6 -
2.3$\times$10$^{-6}$~W/cm, 7 -  5.9$\times$10$^{-6}$~W/cm, 8 -
1.2$\times$10$^{-5}$~W/cm, 9 - 3.7$\times$10$^{-5}$~W/cm;
$f=30$~MHz. Traces are offset vertically for clarity.
 \label{fig3}}
\end{figure*}

The dependences of the real part $\sigma_1$ of the high-frequency conductance, calculated from the SAW attenuation and velocity change, on the reversed
magnetic field 1/$B$ measured at various temperatures from 20~mK to
510~mK are presented in Fig.~\ref{fig3}(a). The dependences
$\sigma_1$(1/$B$) recorded at several SAW
intensities
are plotted in Fig.~\ref{fig3}(b), where
effective SAW power introduced into the sample ranged from
$3.7 \times 10^{-10}$~W/cm to $3.7 \times 10^{-5}$~W/cm.

As seen in Fig.~\ref{fig3}, the Shubnikov - de Haas oscillations are observed at low SAW intensities. These fast oscillations undergo a beating.  At high temperatures  their amplitudes decrease. Moderate increasing of the SAW  power affects the real part of ac conductance $\sigma_1$
in the same way as the temperature rising does, see Fig.~\ref{fig3}(b). However, with further growth of the SAW power, these fast oscillations virtually vanish, and the slow oscillations emerge. The latter dominates at the highest SAW intensities.
The positions of the slow oscillations minima are independent of the SAW frequency.
We assume that the slow oscillations are not distinguishable in Fig.~\ref{fig3}(a) due to the small signal-to-noise ratio in the low-power regime used when we acquired the curves presented in this figure.

\begin{figure}[b]
\centering
\includegraphics[width=\columnwidth]{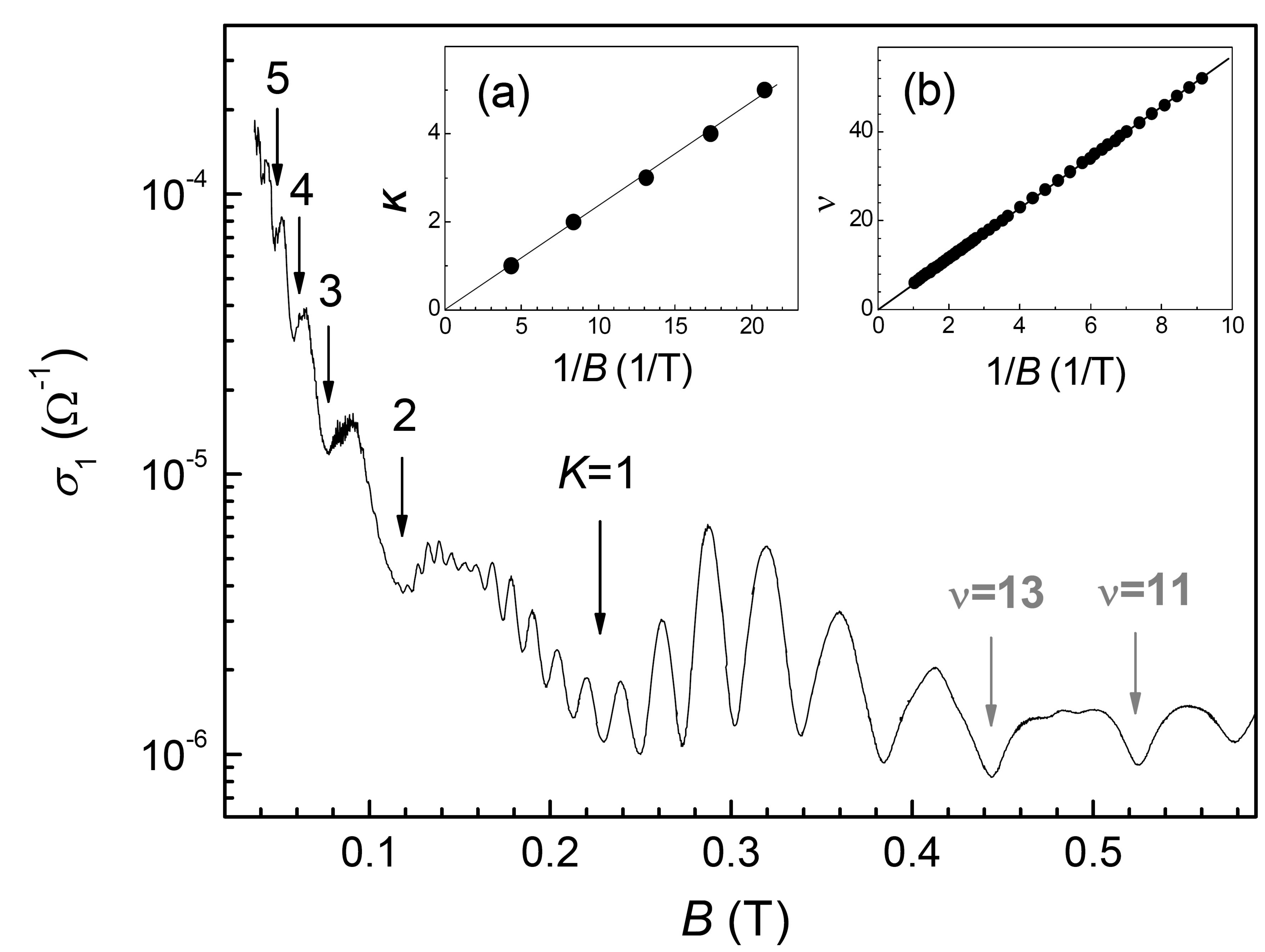}
\caption{Magnetic field dependence of $\sigma_1$ at $f=30$~MHz and
${T \approx 20}$~mK. The SAW power introduced into the sample is $1.2\times10^{-6}$~W/cm.
Inset (a): dependence of the slow oscillations number $K$ on $1/B$. Inset (b): dependence of the filling factors $\nu$ on $1/B$.
 \label{fig4}}
\end{figure}

The structure of the fast and slow oscillations is presented in more detail in Fig.~\ref{fig4}. Here the dependence of $\sigma_1
(B)$ is shown for $f=30$~MHz at 20~mK, the SAW power pushed into the sample was $1.2\times10^{-6}$~W/cm.  This picture
demonstrates the SdH oscillations marked with filling factors $\nu$. In
lower fields $B < 0.4$~T, one can observe a new series of
oscillations denoted by letter $K$.

\section{Discussion}

From the analysis of the  slope of the dependence $\nu(1/B)$ shown in the inset (b) of Fig.~\ref{fig4} we determined
the Fermi energy in the studied WQW as $E_\text{F}\approx$2.5~meV.
The slow oscillations demonstrate a presence of an energy gap $\Delta \ll E_\text{F}$ in the electronic spectrum. We extracted this splitting from the dependence $K(1/B)$ drawn in the inset (a) of Fig.~\ref{fig4}:
 $\Delta =$0.42$\pm$0.02~meV.

In order to explain an origin of this energy splitting, we performed self-consistent calculations of the electrostatic potential and electron wavefunctions. First, the wave functions are calculated in the tight-binding
approach~\cite{Jancu98}.
Then, the electron wave functions are used to calculate the electron density distribution
in the quantum well. Neglecting the dependence of the wave function on the lateral
wave vector, the density is given by the following equation:
\begin{equation}
\label{n}
n(z) = {n_\text{total} \over 4} \sum_{i,s} |\psi_{i,s}(z)|^2,
\end{equation}
where $\psi_{i,\uparrow(\downarrow)}(z)$ is the wave function of a
spin up(down) electron at $i$-th quantum confined level.
The Fermi level lies between 2nd and 3rd levels, so the summation is performed over
the first two subbands.
The value of the total electron density  extracted from our experiment is
$n_{\rm total}=1.4\times10^{11}$cm$^{-2}$.
To compensate the charge inside the WQW and make the structure uncharged,
we assumed that the charge $-n_{\rm total}/2$ is uniformly distributed
in the barriers starting from the
position where the distribution of electron density
$n(z)/n_\text{total}$ drops below $10^{-4}$.
The electrostatic potential
corresponding to the charged QW is found from the numerical solution of
Poisson equation
\begin{equation}
  \phi''(z) = -\frac{4\pi e}{\varepsilon} n(z)\;,
\end{equation}
with the dielectric constant $\varepsilon=12.9$.
Then, we add $\phi(z)$ to the structure potential and compute the
next approximation for the electron wave functions of the levels in the WQW.
The procedure is repeated until the self-consistency of the electron wave functions and
electrostatic potential is reached.

\begin{figure}[t]
\includegraphics[width=0.7\linewidth]{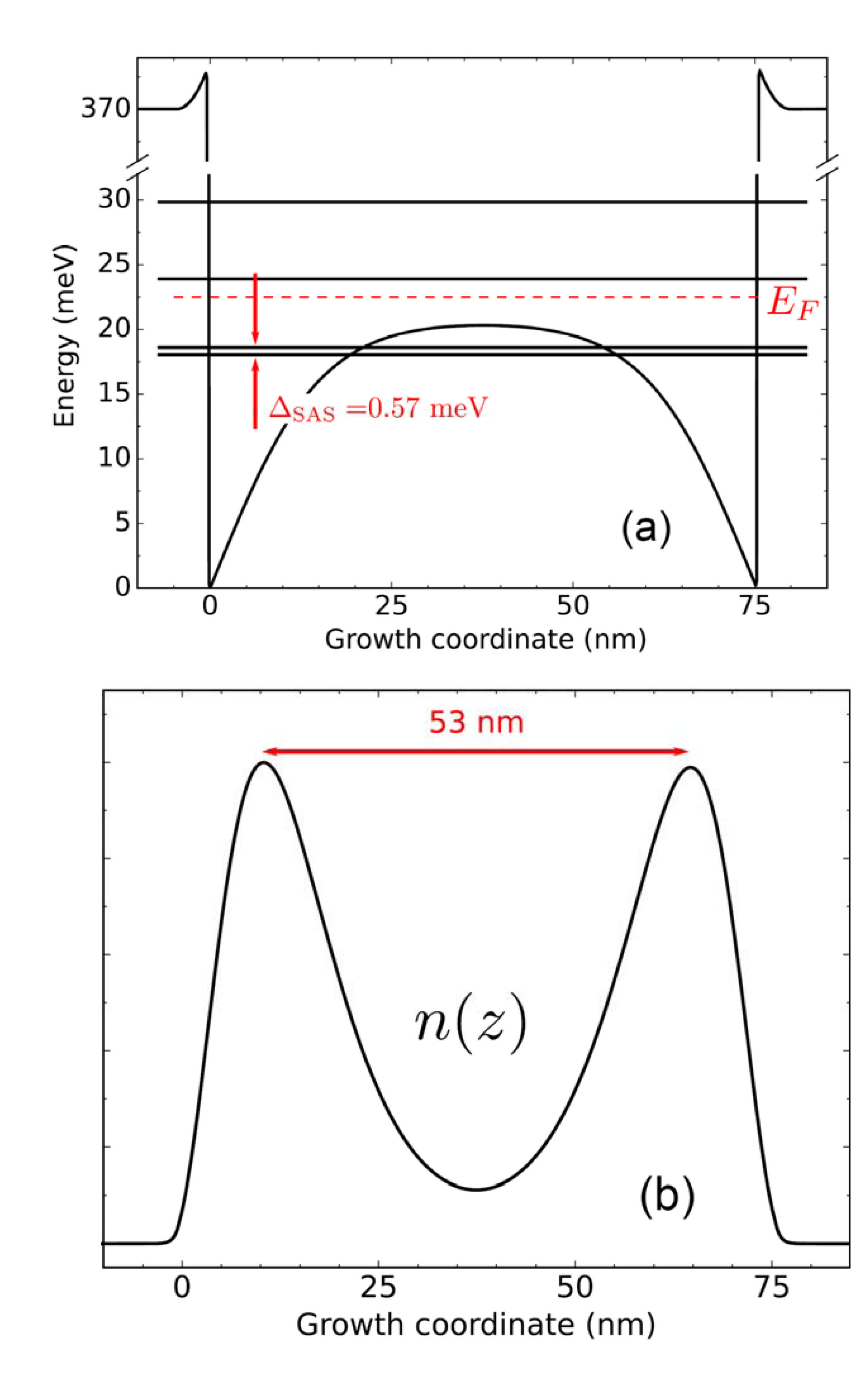}
\caption{(Color online) Self-consistently  calculated energy levels and the heteropotential (a) and  the electron density profile (b).
 \label{fig_calc_energy_wf}}
\end{figure}

The results for the converged potential and the electron density distribution are presented in Fig.~\ref{fig_calc_energy_wf}. The position of the first two levels is close to the local maximum of the heteropotential, Fig.~\ref{fig_calc_energy_wf}(a). This fact makes a convergence of the calculation scheme slow  for our quantum well width and concentration. The electron density profiles shown in  Fig.~\ref{fig_calc_energy_wf}(b) for the two first levels, $|\psi_{1,s}(z)|^2 \approx |\psi_{2,s}(z)|^2$, almost coincide for all $s=\uparrow,\downarrow$.  The distance 53~nm between the density profile maxima agrees with the value for WQWs of the same width~\cite{shayegan_review,Schurova_Galperin}.
The calculated  S-AS splitting  $\Delta_\text{SAS} = 0.57$~meV.

In the triangular quantum wells formed near the structure edges, Fig.~\ref{fig_calc_energy_wf}(a), the  spin-orbit splitting is present which can give rise to the beating pattern in magnetooscillations~\cite{Bychkov_Rashba,Ramvall-experiment}.
Our tight-binding method allows also to estimate the spin splittings of the
two first subbands caused by the quantum confinement and electric field
in the structure~\cite{Pasha17}.
The calculations show that the spin-orbit splitting of the electronic states at the Fermi wavevector is $\Delta_\text{so} \approx 0.01$~meV in the WQW under study.
Since $\Delta_\text{so} \ll \Delta_\text{SAS}$, we conclude that the spin-orbit splitting is negligible at so low carrier density.

The calculated S-AS energy splitting ${\Delta_\text{SAS}=0.57}$~meV is close to the value $\Delta \approx 0.42$~meV determined from the experiment. Therefore we conclude that it is the intersubband scattering that results in slow magnetooscillations of the heated electron gas in the WQW under study.

The conductivity magnetooscillations with  account for both S-AS splitting and  scattering between S and AS subbands are described by the following expression~\cite{I_Dmitriev_review,Raichev_2008}:
\begin{multline}
\label{sigma}
	\sigma_{xx} = {\sigma_0 \over (\omega_c\tau)^2} \\ \times \Biggl[
1 - 4\cos{\left(2\pi{E_\text{F}\over \hbar\omega_c} \right)}\cos{\left( \pi{\Delta_\text{SAS}\over \hbar\omega_c}\right)} \text{e}^{-\pi/\omega_c\tau_q} {X\over \sinh{X}} \\
	+ 2 {\tau\over\tau_\text{SAS}} \cos{\left( 2\pi{\Delta_\text{SAS}\over \hbar\omega_c}\right)} \text{e}^{-2\pi/\omega_c\tau_q}
		\Biggr].
\end{multline}
Here $\sigma_0$ is the conductivity at zero magnetic field, $\tau$ is the transport scattering time which determines the mobility, $\tau_q$ is the quantum scattering time, $\omega_c$ is the cyclotron frequency, and $X=2\pi^2k_\text{B}T/\hbar\omega_c$.
The time $\tau_\text{SAS}$ is the time of elastic scattering between the S and AS subbands.
This expression is valid in moderate magnetic fields where $\text{e}^{-\pi/\omega_c\tau_q} \ll 1$
but $\omega_c\tau \gg 1$, and at weak intersubband scattering, $\tau/\tau_\text{SAS} \ll 1$.
The first oscillating term in Eq.~\eqref{sigma} describes the beating pattern in the Shubnikov-de~Haas oscillations in the two-subband system with close Fermi energies $E_\text{F}\pm \Delta_\text{SAS}/2$. These beatings are damped by heating of the electron gas due to smearing of the Fermi distribution as described by the factor $X/ \sinh{X}$. In contrast, the second oscillating term caused by MISO, being inferior at low temperatures, dominates at high temperatures when $X/ \sinh{X} \ll \text{e}^{-\pi/\omega_c\tau_q}$~\cite{AGTW,Mamani_2008}. Eq.~\eqref{sigma} indicates that the beating frequency to be two times smaller than that for the slow oscillations. Indeed, this is observed in our experiment, Fig.~\ref{fig3}.

We estimated an intensity of intersubband scattering from
the amplitude of MISO. Analysis of the data at the SAW powers $1.2\times10^{-6}$~W/cm and $1.3\times10^{-7}$~W/cm with help of Eq.~\eqref{sigma} yields $\tau/\tau_\text{SAS} = 0.35 \pm 0.05$ and  $\tau_q =4\times 10^{-11}$~s. The value of $\tau_q$ agrees with the quantum scattering time determined for similar WQWs~\cite{tau_q_3_subbands}. The transport scattering time is known from mobility: $\tau=0.9\times 10^{-9}$~s at 0.3~K. This yields $\tau_\text{SAS}=2.6\times 10^{-9}$~s. The intersubband scattering time three times longer than the transport scattering time means that the intersubband scattering in the studied WQW is weaker than the intra-subband scattering but it is strong enough for observation of MISO.

\section{Conclusion}

To conclude, we observed the magneto-intersubband oscillations of the conductivity in a WQW. The oscillations are shown to arise due to elastic intersubband scattering between the S and AS subbands formed due to Coulomb repulsion between the electrons. A tight-binding calculation of the electron states yields the splitting $\Delta_\text{SAS}$ close to the experimentally measured value. Our theoretical description of the magnetooscillations  allowed to determine the quantum and the intersubband scattering times.

\acknowledgments
The authors would like to thank L. Yu. Shchurova for help in calculations, Yu. M. Galperin for discussions and  E. Palm, T.
Murphy, J.-H. Park, and G. Jones for technical assistance.
Partial   support from Presidium of RAS and the  Russian Foundation for Basic Research (project 16-02-0037517)
 is gratefully acknowledged. L.~E.~G. thanks ``BASIS'' foundation. The National High Magnetic Field Laboratory is supported by National Science Foundation Cooperative Agreement No. DMR-1157490 and the State of Florida.
	The work at Princeton was supported by  Gordon and Betty Moore Foundation through the EPiQS initiative Grant GBMF4420, and by the NSF MRSEC Grant DMR-1420541.

\end{document}